\documentclass[aps,prb,twocolumn,showpacs,amsmath,amssymb]{revtex4}

\usepackage{bm}
\usepackage{graphicx}
\usepackage[mathcal]{euscript}
\renewcommand{\vec}[1]{\bm{#1}}
\newcommand{\myfig}[1]{fig#1_colors}  
\DeclareMathOperator{\sign}{sgn}
\begin{document}
\preprint{[DRAFT VERSION]}
\title{Amplitudes for magnon scattering by vortices\\ in two--dimensional weakly
easy--plane ferromagnets} %
\author{Denis D. Sheka} %
\email{Denis_Sheka@univ.kiev.ua} %
\homepage{http://users.univ.kiev.ua/~Denis_Sheka} %
\affiliation{National Taras Shevchenko University of Kiev, 03127 Kiev, Ukraine} %

\author{Ivan A. Yastremsky} %
\affiliation{National Taras Shevchenko University of Kiev, 03127 Kiev, Ukraine} %

\author{Boris A. Ivanov} %
\affiliation{Institute of Magnetism, NASU, 03142 Kiev, Ukraine} %

\author{Gary M. Wysin} %
\affiliation{Department of Physics, Kansas State University, Manhattan, Kansas
66506--2601} %

\author{Franz G. Mertens} %
\affiliation{Physikalisches Institut, Universit\"at Bayreuth, D--95440
Bayreuth, Germany} %
\date{\today}

\begin{abstract}

We study magnon modes in the presence of a vortex in a circular easy--plane
ferromagnet. The problem of vortex--magnon scattering is investigated for
partial modes with different values of the azimuthal quantum number $m$ over a
wide range of wave numbers. The analysis was done by combining analytical and
numerical calculations in the continuum limit with numerical diagonalization of
adequately large discrete systems. The general laws governing vortex--magnon
interactions are established. We give simple physical explanations of the
scattering results: the splitting of doublets for the modes with opposite signs
of $m$, which takes place for the long wavelength limit, is an analogue of the
Zeeman splitting in the effective magnetic field of the vortex. A singular
behavior for the scattering amplitude, $\sigma_m\propto k$, takes place as $k$
diverges; it corresponds to the generalized Levinson theorem and can be
explained by the singular behavior of the effective magnetic field at the
origin.
\end{abstract}



\maketitle


\section{Introduction}
\label{sec:introduction}

It is now firmly established that vortices play an important role in the
condensed matter physics of two--dimensional (2D) systems with continuously
degenerate ground states. In particular, the presence of vortices in 2D
easy--plane (EP) magnets gives rise to the Berezinski\u\i--Kosterlitz--Thouless
phase transition. \cite{Berezinsky72,Kosterlitz73,Kosterlitz74} Vortices play
an essential role in the thermal and dynamical properties of 2D magnets, for a
review see Ref.~\onlinecite{Mertens00}. A vortex signature in dynamical
response functions can be observed experimentally; e.g. translational motion of
vortices leads to a central peak in the dynamical correlation functions.
\cite{Wiesler89} This peak had been predicted predicted both by a vortex gas
theory and by combined Monte Carlo -- spin dynamics simulations.
\cite{Mertens89}

Recently there has been renewed attention to the problem of magnetic vortices
for finite-size magnetic particles, especially their dynamics. It becomes very
important in connection with novel composite magnetic materials such as
magnetic dot arrays.
\cite{Thurn00,Cowburn00,Shinjo00,Sun00,Demokritov01,Erdin02,Cowburn02} These
magnetic dots are submicron--sized islands made from soft magnetic materials on
a nonmagnetic substrate. They are important from a practical standpoint as
high--density magnetic storage devices, \cite{Grimsditch98} and are interesting
as fundamentally new objects in the basic physics of magnetism. The
distribution of magnetization in such a dot is quite nontrivial: when the dot
size is above a critical value, an inhomogeneous state with an out--of plane
magnetic vortex occurs, which is stable due to competition between exchange and
dipole interactions. \cite{Hubert98} It is expected that these nonuniform
states will drastically change the dynamic and static properties of a dot in
comparison with a uniformly magnetized magnetic disk.Recent experiments
\cite{Novosad02a,Park03} verify such properties; in particular, a mode with
anomalously low frequency was detected,\cite{Park03} see also
Refs.~\onlinecite{Guslienko02a,Ivanov02a}.

The general properties of vortex dynamics are intimately connected to the
problem of vortex--magnon interactions. Usually this problem has been studied
numerically for discrete models, mainly for circular samples cut from large
lattice systems.
\cite{Wysin94,Wysin95,Wysin96,Wysin96a,Ivanov98,Wysin01,Kovalev03a,Zagorodny03}
An analytical description of the problem in the framework of the continuum
model has been proposed recently for different 2D magnets.
\cite{Ivanov95g,Ivanov98,Ivanov99,Sheka01,Ivanov02} The most important effect
of the vortex--magnon interaction is an excitation of certain magnon modes due
to vortex motion and vice versa. Because the magnons in the EP ferromagnet have
a gapless dispersion law, a possible Larmor dynamics of the vortex center is
strongly coupled with a magnon cloud; \cite{Kovalev03a} therefore the
corresponding motion has a non--Newtonian form. \cite{Ivanov98}

In this paper we consider the magnon modes which exist in a 2D Heisenberg EP
ferromagnet in the presence of a vortex. We apply different, sometimes new,
methods of analytical and numerical investigation, in order to extend the
research of Ref.~\onlinecite{Ivanov98}, presenting a wider range of results for
the magnon scattering amplitude. In Sec.~\ref{sec:model} we demonstrate that
the vortex acts on magnons in two ways. First of all, it provides for coupling
between different directions of magnetization precession; the magnon modes are
described by a ``generalized'' Schr\"{o}dinger equation. Secondly, the problem
naturally possesses an effective magnetic field, whose global properties are
caused by the soliton topological charge. The scattering problem is treated
both numerically and analytically for a wide range of wave numbers. The
numerical study is carried out using two different approaches: solving the
eigenvalue problem for the continuum limit (in the weak EP anisotropy limit),
and extracting the scattering data from numerical diagonalization of discrete
systems, see Sec.~\ref{sec:scat-num}. The analytical study of the scattering
problem is developed (Sec.~\ref{sec:scat-analyt}) using both the long-- and
short--wavelength approximations. In contrast to the previous work
\cite{Ivanov98} we describe analytically the splitting phenomenon of the
doublets of magnon modes with opposite signs of the azimuthal quantum number,
and give a physical picture of this effect: an effective magnetic field causes
the Zeeman splitting of the magnon levels, see Sec.~\ref{sec:k<<1}. The
singular behavior of the scattering amplitude is predicted in
Sec.~\ref{sec:k>>1} for the short wavelength limit. This feature is caused by
the specific singular effective magnetic field at the origin (vortex core);
this study verifies the generalized version of the Levinson theorem, which we
have established recently in Ref.~\onlinecite{Sheka03} for potentials with
inverse square singularities.


\section{Model and magnon modes} %
\label{sec:model}

We consider the classical 2D Heisenberg ferromagnet (FM) with the Hamiltonian
\begin{equation} \label{eq:Hamiltonian}
\mathcal{H} = -J\sum_{\left(\vec{n},\vec{n}'\right)}\left[
\vec{S}_{\vec{n}}\cdot\vec{S}_{\vec{n}'} -(1-\lambda){S}_{\vec{n}}^z
{S}_{\vec{n}'}^z \right],
\end{equation}
where the spins $\vec{S}_{\vec{n}}$ are classical vectors on a square lattice
with the lattice constant $a$. Here $\left(\vec{n},\vec{n}'\right)$ denotes
nearest--neighbor lattice sites, $J>0$ is the exchange integral, and
$\lambda\in[0,1)$ describes easy--plane anisotropy.

We consider the continuum dynamics of the Heisenberg ferromagnet, which is
adequate for model \eqref{eq:Hamiltonian} in the small--anisotropy case
($1-\lambda\ll1$). In a continuum limit the dynamics of the FM is described by
Landau--Lifshitz equations for the normalized magnetization,
\begin{equation} \label{eq:m}
\vec{m}(\vec{r},t) = \frac{\vec{S}(\vec{r},t)}{S} =
\Bigl(\sin\theta\cos\phi;\sin\theta\sin\phi;\cos\theta\Bigr).
\end{equation}
The equations of motion result from the Lagrangian
\begin{equation} \label{eq:Lagrangian}
L = \frac{S}{a^2}\int \text{d}^2x(1-\cos\theta)\frac{\partial\phi}{\partial t}
- E[\theta,\phi],
\end{equation}
with the energy functional
\begin{equation} \label{eq:Energy}
E[\theta,\phi] = \frac{JS^2}{2}\!\!\int\!\! \text{d}^2x\! \left[\left(\nabla
\theta \right) ^2+\left( \nabla \phi \right) ^2\sin ^2\theta +
\frac{\cos^2\theta}{r_v^2}\right].
\end{equation}
Here $r_v=\frac a2\sqrt{\lambda/(1-\lambda)}$ is the characteristic length
scale (``magnetic length''). One should note that the strength of the weak
easy-plane anisotropy ($\lambda\approx 1$), determines the magnetic length
scale $r_v$; in this continuum analysis, results will depend on lengths scaled
by $r_v$, and have no explicit dependence on the anisotropy strength.

The ground state of the magnet is continuously degenerate and isotropic within
the easy plane (EP). The simplest elementary linear excitations of EP FMs that
arise on the homogeneous background are the magnons belonging to a continuous
spectrum. They have the form of elliptically polarized waves; if one chooses
the homogenous spin distribution along the $x$--axis, then the spin wave takes
the form:
\begin{equation} \label{eq:spin-wave}
m_x=1,\quad m_z+{i} m_y \cdot \frac{kr_v}{\sqrt{1+k^2r_v^2}}\propto \exp({i}
kx-{i} \omega t).
\end{equation}
The dispersion law has the gapless form:
\begin{equation}\label{eq:dispersion_law} \omega(\vec{k}) =
ck\sqrt{1+k^2r_v^2},
\end{equation}
where $c=2aJS\sqrt{1-\lambda}$ is the characteristic magnon speed, $\vec{k}$ is
the magnon wave vector, and $k=|\vec{k}|$ is its magnitude.

The simplest nonlinear excitation in the system is an out--of--plane (OP)
vortex \cite{Gouvea89a,Kosevich90}
\begin{equation} \label{eq:vortex}
\begin{split}
&\phi\equiv\phi_0=\varphi _0 + q \chi,\quad \theta=\theta_0(\rho),\\
&\theta_0(0)=\frac{1-p}{2}\pi,\quad \theta_0(\infty)=\pi/2,
\end{split}
\end{equation}
where $\varphi_0$ is an arbitrary angle due to the EP symmetry,
$\rho\equiv|\vec{r}|/r_v$ and $\chi$ are dimensionless polar coordinates in the
plane of the magnet, the vorticity $q\in\mathbb{Z}$ plays the role of a $\pi_1$
topological charge, and the polarization $p=\pm1$ is connected with a $\pi_2$
topological charge (the Pontryagin index):
\begin{equation} \label{eq:Pontryagin-index}
Q=\frac{1}{4\pi}\int\mathcal{Q}\;\text{d}^2x,\qquad
\mathcal{Q}=\frac12\epsilon_{ij}\
\vec{m}\cdot\left(\partial_i\vec{m}\times\partial_j\vec{m}\right).
\end{equation}
We term $\mathcal{Q}$ the gyrocoupling density, following \citet{Thiele73}; it
has a sense as the density of the topological charge, see \citet{Nicos91}. For
the vortex configuration \eqref{eq:vortex} the gyrocoupling density can be
represented as
\begin{equation} \label{eq:Q}
\mathcal{Q}=\frac{q\sin\theta_0\cdot\theta_0^\prime}{\rho}, \end{equation}
hence the Pontryagin index takes on half--integer or integer values $Q=qp/2$.
Note that the presence of a nontrivial $\pi_2$--topological charge directly
results in the gyrotropical dynamics of the vortex, which conserves the
gyrovector $\vec{G}=Q\cdot2\pi\hslash S a^{-2}\vec{e}_z$.

The function $\theta_0$ is the solution of an ordinary differential equation,
which can only be solved numerically. \cite{Kosevich90,Bar'yakhtar93} Without
an external magnetic field two oppositely polarized vortices are energetically
equivalent; for definiteness we set $p=+1$.

To analyze magnons on the vortex background, we use a formalism and set of
coordinates developed in Ref.~\onlinecite{Wysin95}, setting up the problem in
terms of local Cartesian spin components. The unperturbed spins of the static
vortex structure, $\vec{m}_0$, define \emph{local} polar axes $\vec{e}_3$,
different at every site, specifically, $\vec{S}_0(\vec{r},t)=S\vec{e}_3$.

It is to be understood that these axes depend on the site chosen. The magnetic
fluctuations occur perpendicular to these local axes, suggesting the definition
of other axes, $\vec{e}_2$ being chosen along the direction of $\vec{e}_z
\times \vec{e}_3$, and $\vec{e}_1 = \vec{e}_2 \times \vec{e}_3$, to complete
the mutually perpendicular set. One supposes that a dynamically fluctuating
spin has small deviations along the $\vec{e}_1$ and $\vec{e}_2$ axes so that a
spin is written as
\begin{equation} \label{eq:m1m2-def} %
\vec{S}(\vec{r},t) = S\left(\vec{e}_3 +m_1 \vec{e}_1+m_2\vec{e}_2\right).
\end{equation}
The fields $m_1$ and $m_2$ have a simple physical significance, which can be
seen if a given spin is supposed to have small deviations $\varphi$ and
$\vartheta$ away from the vortex structure, determined by azimuthal and polar
spherical angles, $\phi_0$ and $\theta_0$. We write
\begin{equation} \label{eq:var-def} %
\begin{split}
\frac{\vec{S}(\vec{r},t)}{S}&= \cos(\theta_0+\vartheta)\vec{e}_z +
\sin(\theta_0+\vartheta)\\ &\times \left[ \cos(\phi_0+\varphi)
\vec{e}_x+\sin(\phi_0+\varphi)\vec{e}_y \right].
\end{split}
\end{equation}
Linearizing in $\varphi$ and $\vartheta$, and using the definitions of $\{
\vec{e}_1, \vec{e}_2, \vec{e}_3 \}$, comparison of Eqs.~\eqref{eq:m1m2-def} and
\eqref{eq:var-def} shows that
\begin{equation} \label{eq:m12-via-var} %
m_1=\vartheta, \quad m_2= \varphi \sin\theta_0.
\end{equation}
Thus, the $m_1$ field measures spin rotations moving towards the polar
($\vec{e}_z$) axis and the $m_2$ field measures spin rotations projected onto
the $xy$--plane. In the absence of the vortex, we have $\theta_0=\pi/2$,
$\phi_0=0$, and such oscillations correspond to the free magnons in the form
\eqref{eq:spin-wave}.

The linearized equations for $m_1$ and $m_2$ can be described by a single
complex--valued function $\psi(\vec{r},t) = m_1 + {i} m_2$, which obeys the
differential equation
\begin{equation} \label{eq:Gen-Schroedinger}
{i}\partial_\tau \psi=H\psi+W\psi^*,\qquad %
H=\left(-{i}\vec{\nabla}-\vec{A}\right)^2 + U,
\end{equation}
with the ``potentials''
\begin{subequations} \label{eq:potentials}
\begin{eqnarray} \label{eq:Gen-U}
U(\rho)&=&\frac12\sin^2\theta_0\left(1-\frac{q^2}{\rho^2}\right) -
\cos^2\theta_0 -
\frac{{\theta'_0}^2}{2},\\
\label{eq:Gen-W} %
W(\rho)&=&\frac12\sin^2\theta_0\left(1-\frac{q^2}{\rho^2}\right) +
\frac{{\theta'_0}^2}{2},\\
\label{eq:Gen-A} %
\vec{A}(\rho) &=& -\frac{q\cos\theta_0}{\rho}\cdot\vec{e}_\chi.
\end{eqnarray}
\end{subequations}
Here we use the dimensionless coordinate variable $\rho=|\vec{r}|/r_v$,
dimensionless time variable $\tau = t\cdot c/r_v$, and the operator
$\vec{\nabla}=r_v\partial_{\vec{r}}$; prime denotes $\text{d}/\text{d}\rho$.

Let us note that the vector $\vec{A}$ acts in the Schr\"{o}dinger--like
operator $H$ in the same way as the vector--potential acts in the Hamiltonian
of a charged particle. Then it is possible to conclude that there is an
effective magnetic flux density
\begin{equation} \label{eq:B-eff}
\vec{B}=\vec{\nabla}\times\vec{A}=\vec{e}_z\cdot\frac{q\sin\theta_0\cdot
\theta'_0}{\rho}.
\end{equation}
Note that the effective magnetic flux density can easily be rewritten through
the gyrocoupling density \eqref{eq:Q} as $\vec{B}=\mathcal{Q}\cdot\vec{e}_z$.
Therefore the total flux is determined by the nontrivial $\pi_2$ topology of
the vortex configuration. On first view when exploiting this analogy it is
possible to look for the Aharonov--Bohm phenomenon for the scattering problem,
because this magnetic flux density is localized in the region of the vortex
core. However, one can see that the total magnetic flux
\begin{equation} \label{eq:flux}
\Phi=\int B_z \text{d}^2x = 4\pi Q = qp\Phi_0
\end{equation}
is an integer multiple of the flux quantum $\Phi_0=2\pi$, so there is no
Aharonov--Bohm scattering picture for the system.

A differential equation like \eqref{eq:Gen-Schroedinger} is not a unique
property of the vortex--magnon problem in the EP FM only. It appears for
different kinds of anisotropy: it describes magnon modes on the soliton
background in the easy--axis \cite{Sheka01} and isotropic magnets
\cite{Ivanov99}. Note that for the specific case of an isotropic system with an
exact analytical soliton solution of the Belavin--Polyakov type, the potential
$W$ disappears, so the magnon modes satisfy the usual Schr\"{o}dinger like
equation ${i}\partial_\tau \psi = H\psi$, which describes, e.g., the quantum
mechanical states for a charged particle in the axially symmetric potential
$U(\rho)$ under the action of an external magnetic field with a vector
potential $\vec{A}$.

For the anisotropic case, when $W\neq0$, the problem
\eqref{eq:Gen-Schroedinger} has important unusual properties, which are absent
for the Belavin--Polyakov case. More generally, there appear properties which
are forbidden for the usual quantum mechanics. In particular, an effective
discrete Hamiltonian of the system is not necessarily Hermitian; in
Refs.~\onlinecite{Wysin95,Wysin96,Wysin01} some constructive methods were
elaborated to avoid these problems. Nevertheless we will discuss the features
of Eq.~\eqref{eq:Gen-Schroedinger} in order to understand why the standard
quantum mechanical intuition could fail.

The standard quantum mechanical equation ${i}\partial_\tau \psi = H\psi$ allows
the conservation law $\partial_\tau|\psi|^2=-\vec{\nabla}\cdot\vec{j}$ for the
current
\begin{equation} \label{eq:current}
\vec{j} = {i} \left(\psi\vec{\nabla}\psi^* - \psi^\star\vec{\nabla}\psi\right)
+ 2|\psi|^2\vec{A}.
\end{equation}
The generalized Schr\"{o}dinger--like equation \eqref{eq:Gen-Schroedinger} with
$W\neq0$ violates this conservation law, namely:
\begin{equation} \label{eq:div-j}
\partial_\tau|\psi|^2=-\vec{\nabla}\cdot\vec{j}-{i} W\left({\psi^*}^2 -
\psi^2\right).
\end{equation}

Nonconservation of probability density has posed some problems in the passage
from standard quantum mechanics to old pre--Feynmann quantum electrodynamics.
The reason is that Eq.~\eqref{eq:Gen-Schroedinger} is formulated neither for a
Hermitian, nor a linear operator; the last statement is due to the broken
symmetry under the rescaling $\psi\to\lambda\psi$ with $\lambda\in\mathbb{C}$.
There exists an analogy with relativistic theory: there can appear solutions
with positive and negative energy in the passage from the Klein--Gordon to
Dirac equation. In fact, our problem has the same origin. Let us reformulate
the problem \eqref{eq:Gen-Schroedinger} as an equation second order in time.
One can calculate that the Klein--Gordon--like equation
\begin{equation} \label{eq:KG}
-\partial_{\tau\tau}\psi = \left(H^2-W^2\right)\psi
\end{equation}
is valid far from the vortex center. What is important is that
Eq.~\eqref{eq:KG} contains a Hermitian operator (similar arguments were used in
Ref.~\onlinecite{Wysin01}). Therefore, the eigenvalue problem (EVP) for
$\omega^2$, not for $\omega$, is more appropriate for this system; the only
problem is to separate solutions with positive and negative $\omega$. Note,
that there appear $4^{\rm th}$ order operators with respect to the space
coordinate, which causes the presence of master and slave functions in the
solution, see below.

Such a problem, as well as a problem with nonconserved number of particles
(probability amplitude), appears in the theory of a weakly nonideal Bose gas.
It results, in fact, in the separation of positive and negative energy
solutions under u--v Bogolyubov transformations. \cite{LandauIX}

Following this scheme we need to generalize the u--v transformation to the
nonhomogeneous case.

We apply the partial--wave expansion, using the \emph{ansatz} \cite{Ivanov98}
\begin{equation} \label{eq:Psi-via-u&v}
\begin{split}
\psi(\vec{r},t)&=\sum_{\alpha} \left[u_\alpha(\rho)e^{{i}\Phi_\alpha} +
v_\alpha(\rho)e^{-{i}\Phi_\alpha}\right],\\
\Phi_\alpha(\chi,t) &= m\chi-\omega_\alpha t + \eta_m = m\chi-\varOmega_\alpha
\tau + \eta_m,
\end{split}
\end{equation}
where $\alpha=(k,m)$ is a full set of eigenvalues, $m\in\mathbb{Z}$ being
azimuthal quantum numbers, the $\eta_m$ are arbitrary phases, and
$\varOmega=\omega r_v/c$ are dimensionless frequencies. This expansion leads to
the following EVP for the radial eigenfunctions $u$ and $v$ (the index $\alpha$
will be omitted in the following):
\begin{equation} \label{eq:EVP-u&v} \textsf{H}\bm{\bigl|\varPsi\bigr>} =
\varOmega\bm{\bigl|\varPsi\bigr>}, \; \textsf{H}=
\begin{Vmatrix}H_+ & W \\ -W & -H_- \\ \end{Vmatrix},\;
\bm{\bigl|\varPsi\bigr>} =
\begin{Vmatrix} u \\ v \end{Vmatrix}.
\end{equation}
Here $H_\pm = -\nabla_\rho^2+\mathcal{U}_0+1/2 \pm V$ is the 2D radial
Schr\"{o}dinger--like operator with the ``potentials''
\begin{align}
&\mathcal{U}_0(\rho)= U(\rho)+\vec{A}^2+\frac{m^2}{\rho^2}-\frac12 \nonumber \\
\label{eq:U} %
&= \frac{q^2+m^2}{\rho^2} -\frac{3q^2\sin^2\theta_0}{2\rho^2}
-\frac{3\cos^2\theta_0}{2} -
\frac{{\theta'_0}^2}{2},\\
\label{eq:V} %
&V(\rho) =-\frac{2m\left(\vec{A}\cdot\vec{e}_\chi\right)}{\rho}
=\frac{2qm\cos\theta_0}{\rho^2},
\end{align}
$\nabla_\rho^2=\text{d}^2/\text{d}\rho^2+(1/\rho)\text{d}/\text{d}\rho$ is the
radial Laplace operator. In spite of the fact that the EVP \eqref{eq:EVP-u&v}
is formulated for the Schr\"{o}dinger operators $H_\pm$, this EVP is different
in principle from the usual set of coupled Schr\"{o}dinger equations, which is
widely used, e.g. for the description of multichannel scattering.
\cite{Newton82} The reason is that the matrix Hamiltonian $\textsf{H}$ is not
Hermitian for the standard metric, for details see Ref.~\onlinecite{Wysin01}.
To avoid this problem we introduce a corresponding bra--vector by the
definition
\begin{equation} \label{eq:bra-vector}
\bm{\bigl<\varPsi\bigl|} = \bigl\| u\;; -v \bigr\|.
\end{equation}
The Hilbert space for the $\vec{\varPsi}$--function has an indefinite metric,
\begin{equation} \label{eq:norm}
\bm{\bigl<\varPsi\bigr|\varPsi\bigr>} = (u|u)-(v|v),
\end{equation}
where $(u|v)=\int_0^\infty u(\rho)v(\rho)\rho \text{d}\rho$ is the standard
scalar product. By introducing such a Hermitian product, it is possible to
define the standard energy functional, cp. Ref.~\onlinecite{Sheka01}:
\begin{equation} \label{eq:E-funct}
\mathcal{E}[u,v]=\bm{\bigl<\varPsi\bigl|\textsf{H}\bigr|\varPsi\bigr>} =
(u|H_+|u)+2(u|W|v)+(v|H_-|v).
\end{equation}

Let us mention that Eq.~\eqref{eq:EVP-u&v} is invariant under the conjugations
$\varOmega\to-\varOmega$, $m\to-m$, and $u\leftrightarrow v$. In a classical
theory we can choose either sign of the frequency; but in order to make contact
with quantum theory, with a positive frequency and energy
$\mathcal{E}_k=\hslash\omega_k$, we will discuss the case $\varOmega>0$
($\omega>0$) only. Thus there are two different equations for the opposite
signs of $m$. However, in the limiting case of the ``zero modes'' with
$\varOmega=0$, the system again is invariant under conjugations $m\to-m$. For
example, one of the zero modes, the so--called translational mode with $m=+1$,
has the form
\begin{equation} \label{eq:zero-mode-m=+1}
u_{m=+1} = \frac{\sin\theta_0}{\rho} - \theta'_0,\qquad v_{m=+1} =
\frac{\sin\theta_0}{\rho} + \theta'_0,
\end{equation}
which describes the position shift of the soliton. Because of the degeneration
of the EVP at $\varOmega=0$, it leads to the existence of a zero mode with
$m=-1$; the eigenfunction of this mode can be expressed just from
\eqref{eq:zero-mode-m=+1} under the conjugation $u\leftrightarrow v$. We use
here notations for mode indices as in Refs.~\onlinecite{Ivanov99,Ivanov02};
note that the mode with $m=+1$ corresponds in our notation to the mode with
$m=-1$ in the notations of Refs.~\onlinecite{Ivanov98,Sheka01}.

It should be stressed that the picture is quite different for the special case
$W=0$, which corresponds to the isotropic magnet. \cite{Ivanov95g,Ivanov99}
Here we have two uncoupled equations for the functions $u$ and $v$. One of the
equations (for the eigenfunction $v$) has the negative eigenvalue $-\varOmega$,
from which it necessarily results that $v\equiv0$. In this special case the
zero modes \eqref{eq:zero-mode-m=+1} have the form $u_{+1}\propto \theta_0'$
and $v_{+1}=0$. Therefore the zero mode with $m=-1$ cannot be obtained by the
simple conjugation. It explains the difference between the collective dynamics
of the soliton in isotropic magnets, where it is enough to take into account
only the mode with $m=+1$, and the EP FM, where translational modes with $m=-1$
and $m=+1$ must be taken into account. Nevertheless, the roles of the modes
with $m=-1$ and $m=+1$ are not equal, for details see Sec.~\ref{sec:Levinson}.


\section{Scattering problem: numerical results}
\label{sec:scat-num}


\subsection{Continuum approach}
\label{sec:scat-cont}

We intend to describe the scattering of magnons by a vortex. However the EVP
\eqref{eq:EVP-u&v} is not adjusted for the scattering problem, because it does
not provide the asymptotic independence of the equations at infinity. To solve
the problem it is convenient to make a unitary transformation of the
eigenvector $\bm{\bigl|\varPsi\bigr>}$,
\begin{equation} \label{eq:tilde-Psi}
\bm{\bigl|\widetilde{\varPsi}\bigr>} = \textsf{A} \bm{\bigl|\varPsi\bigr>},
\quad \textsf{A}=
\begin{Vmatrix}
\cos\varepsilon & -\sin\varepsilon \\
\sin\varepsilon & \cos\varepsilon \\
\end{Vmatrix},
\; \bm{\bigl|\widetilde{\varPsi}\bigr>} = \begin{Vmatrix} \tilde{u} \\
\tilde{v}
\end{Vmatrix}.
\end{equation}
The angle $\varepsilon$ of this unitary transformation is defined by the
expression
\begin{equation} \label{eq:eps}
\tan2\varepsilon = \frac{1}{2\varOmega}.
\end{equation}
Then we obtain the following partial differential equation for the function
$\bm{\bigl|\widetilde{\varPsi}\bigr>} $:
\begin{subequations} \label{eq:PDE4tilde-Psi+tilde-H}
\begin{align} \label{eq:PDE4tilde-Psi}%
&\widetilde{\textsf{H}} \bm{\bigl|\widetilde{\varPsi}\bigr>} = \bm{\Lambda}
\bm{\bigl|\widetilde{\varPsi}\bigr>}, \; \widetilde{\textsf{H}} = \textsf{H}_0
+ \widetilde{\textsf{V}}, \;\bm{\Lambda}=
\text{diag}\left(\kappa^2;\varkappa^2\right),\\
\label{eq:H0} %
&\textsf{H}_0 =\text{diag}\left(\mathcal{H}_0;-\mathcal{H}_0\right),\quad
\mathcal{H}_0 =  -\nabla_\rho^2+\mathcal{U}_0,\\
\label{eq:tilde-V} %
&\widetilde{\textsf{V}} = \left[V+\bm{g}\cdot\left(W-1/2\right)\right]
\textsf{A}^{-2},
\end{align}
\end{subequations}
where $\bm{g}= \bigl\|\begin {smallmatrix}0&1\\-1&0\end{smallmatrix}\bigr\|$ is
a metric spinor, the dimensionless wave number is $\kappa = kr_v$, and
$\varkappa=\sqrt{\kappa^2+1}$.

First let us consider the magnon spectrum in the absence of a vortex (free
fields). Without a vortex ($q=0$, $\theta_0=\pi/2$),
Eqs.~\eqref{eq:PDE4tilde-Psi+tilde-H} are uncoupled, which results in free
magnons,
\begin{equation} \label{eq:modes_free}
\begin{split}
\tilde{u}_m(\rho) &\propto J_{|m|}(\kappa \rho) \underset{\kappa
\rho\gg1}{\sim} \sqrt{\frac{2}{\pi\kappa\rho}}\cdot\cos\left(\kappa \rho -
\frac{|m|\pi}{2} -\frac{\pi}{4}\right),\\
\tilde{v}_m(\rho) &=0,
\end{split}
\end{equation}
where $J_m$ are Bessel functions. The free modes $\tilde{u}_m$ play the role of
the partial cylinder waves of a plane spin wave
\begin{equation} \label{eq:plain_wave}
\exp\left({i}\vec{k}\cdot\vec{r}-{i}\omega t\right) = \sum_{m=-\infty}^\infty
{i}^mJ_m(\kappa \rho)e^{{i} m\chi-{i} \omega t}.
\end{equation}

To describe magnon solutions in the presence of a vortex, one should note that
far from the vortex center the potential $\widetilde{\textsf{V}}$ tends to
zero, so the Eqs.~\eqref{eq:PDE4tilde-Psi} become uncoupled,
\begin{equation} \label{eq:PDE-as}
\left( \nabla_\rho^2 + \kappa^2 \right) \tilde{u} =0,\; %
\left( \nabla_\rho^2 - \varkappa^2 \right) \tilde{v} =0,\; %
\rho\gg\max\left(1;\frac1\kappa;\frac1\varkappa\right)
\end{equation}
with asymptotically independent solutions:
\begin{subequations} \label{eq:u&v4x>>1}
\begin{align}
\tilde{u}_m(\rho) & \sim \frac{C_1}{\sqrt{\rho}}\cdot e^{{i}\kappa\rho} +
\frac{C_2}{\sqrt{\rho}}\cdot e^{-{i}\kappa\rho}\nonumber\\
\label{eq:u4x>>1} %
&\propto \frac{1}{\sqrt{\rho}}\cos\left(\kappa
\rho - \frac{|m|\pi}{2} -\frac{\pi}{4} + \delta_m \right),\\
\label{eq:v4x>>1} %
\tilde{v}_m(\rho) & \sim \frac{C_3}{\sqrt{\rho}}\cdot e^{\varkappa\rho} +
\frac{C_4}{\sqrt{\rho}}\cdot e^{-\varkappa\rho}.
\end{align}
\end{subequations}

The scattering results in the quantity $\delta_m\equiv\delta_m(\kappa)$; it can
be interpreted as the scattering phase shift, determining the intensity of the
magnon scattering due to the presence of the vortex. Sometimes it is useful to
introduce the scattering amplitude, $\sigma_m=-\tan\delta_m$. Using this
notation, the oscillatory solution \eqref{eq:u4x>>1} can be rewritten in the
following form
\begin{equation} \label{eq:u-via-J&Y}
\tag{\ref{eq:u4x>>1}$'$} %
\tilde{u}_m(\rho) \propto J_{|m|}(\kappa \rho) + \sigma_m \cdot Y_{|m|}(\kappa
\rho).
\end{equation}
where $Y_{|m|}$ are Neumann functions. Let us stress that the solution
\eqref{eq:u-via-J&Y} is valid only in the sense of the asymptotic form
\eqref{eq:u4x>>1}. As it follows from Eq.~\eqref{eq:v4x>>1}, the function
$\tilde{v}$ has an exponential behavior,
\begin{equation} \label{eq:v-via-K&I}
\tag{\ref{eq:v4x>>1}$'$} %
\tilde{v}_m(\rho) \propto K_{|m|}(\varkappa \rho) + \gamma_m \cdot
I_{|m|}(\varkappa \rho)\propto \frac{e^{-\varkappa\rho}}{\sqrt{\rho}} +
\gamma_m\cdot \frac{e^{\varkappa\rho}}{\sqrt{\rho}},
\end{equation}
where $K_{|m|}$ and $I_{|m|}$ are MacDonald and modified Bessel functions,
respectively; at the same time $\tilde{u}$ yields oscillatory solutions.
Naturally, the real modes have an oscillatory form here; we will use this fact
below for the numerical analysis. It means that the function $\tilde{u}$
becomes a master function in the Eq.~\eqref{eq:PDE4tilde-Psi}, while
$\tilde{v}$ is a slave (note, that we choose $\varOmega>0$). This mirrors the
difference between Eq.~\eqref{eq:PDE4tilde-Psi} and a usual set of
Schr\"{o}dinger equations.

The scattering amplitude, or, equivalently, the phase shift, contains all
information about the scattering processes. In particular, the general solution
of the scattering problem for a plane wave can be expressed in the form, cp.
Eq.~\eqref{eq:spin-wave}
\begin{subequations} \label{eq:G4plane-wave}
\begin{equation} \label{eq:G4plane-wave1}
m_2-{i} m_1\cdot \frac{kr_v}{\sqrt{1+k^2r_v^2}} \propto
e^{{i}\vec{k}\cdot\vec{r}-{i}\omega t} + \mathcal{F}(\chi)\cdot
\frac{e^{{i}\kappa \rho-{i}\omega t}}{\sqrt{\rho}},
\end{equation}
where the scattering function has the form \cite{Ivanov99}
\begin{equation} \label{eq:G4plane-wave2}
\mathcal{F}(\chi) = \frac{\exp(-{i}\pi/4)}{\sqrt{2\pi\kappa}}\cdot
\sum_{m=-\infty}^{\infty} \left(e^{2{i}\delta_m}-1\right)\cdot e^{{i} m\chi}.
\end{equation}
\end{subequations}

The total scattering cross section is given by the expression
\begin{equation*}
\varrho = \int_0^{2\pi}\!|\mathcal{F}|^2\text{d}\chi =
\sum_{m=-\infty}^\infty\! \varrho_m\;,
\end{equation*}
where $\varrho_m = (4/k)\sin^2\delta_m$ are the partial scattering cross
sections.

Let us switch to the numerical solution of the scattering problem in the
continuum approach. The differential problem to be integrated consists of
Eq.~\eqref{eq:PDE4tilde-Psi+tilde-H} and asymptotic conditions at the center of
the vortex and at infinity:
\begin{subequations} \label{eq:PDE4num}
\begin{align} \label{eq:PDE4num(1)} %
\widetilde{\textsf{H}} \bm{\bigl|\widetilde{\varPsi}\bigr>} &= \bm{\Lambda}
\bm{\bigl|\widetilde{\varPsi}\bigr>},\\
\label{eq:PDE4num(2)} %
\bm{\bigl|\widetilde{\varPsi}\bigr>} & \sim \textsf{A}\cdot
\begin{Vmatrix}\epsilon_m\cdot\rho^{|m+1|},\\ %
\rho^{|m-1|}\end{Vmatrix},\qquad \text{when}\;\rho\ll1\\
\nonumber %
\bm{\bigl|\widetilde{\varPsi}\bigr>} & \sim
\begin{Vmatrix}
J_{|m|}(\kappa\rho)+\sigma_m\cdot Y_{|m|}(\kappa\rho) \\ %
K_{|m|}(\varkappa\rho)
\end{Vmatrix},\\
\label{eq:PDE4num(3)} %
&\qquad\text{when}\;\rho\gg\max\left(1;\frac1\kappa; \frac1\varkappa\right).
\end{align}
\end{subequations}
The presence of the matrix $\textsf{A}$ in the condition \eqref{eq:PDE4num(2)}
means that the functions $\tilde{u}$ and $\tilde{v}$ are not asymptotically
independent even in the lowest approximation. In the next approximation there
appears an additional ``interaction'' between $\tilde{u}$ and $\tilde{v}$,
which is realized in the nonunit factor $\epsilon_m$; its value cannot be found
through this asymptotic expansion.

We use the one--parameter shooting method, solving Eqs.~\eqref{eq:PDE4num}, as
described in Ref.~\onlinecite{Ivanov96,Sheka01}. Choosing the shooting
parameter $\epsilon_m$, we ``kill'' the growing exponent for the function
$\tilde{v}_m$ in \eqref{eq:v-via-K&I}, where the coefficient $\gamma_m$ should
be equal to zero; as a result we have obtained a well--pronounced exponential
decay for $\tilde{v}_m\propto K_{|m|}(\varkappa \rho)$, and oscillating
solutions for $\tilde{u}_m$. The scattering amplitude was found from these data
by comparison with the asymptotes \eqref{eq:PDE4num(3)}. The results are
discussed in Sec.~\ref{sec:scat-disc}.


\subsection{Discrete approach}
\label{sec:scat-discrete}

In the discrete lattice approach, the small amplitude spin fluctuation modes in
the presence of a vortex at the center of a finite circular system of radius
$R$ are found. The spins occupy sites on a square lattice. We use the formalism
and set of local coordinates as described in Sec.~\ref{sec:model} for the
continuum model. Similar to the continuum expression \eqref{eq:m1m2-def}, we
describe the dynamically fluctuating spin on lattice site $\vec{n}$ as
\begin{equation} \label{eq:S-fluct} %
\tag{\ref{eq:m1m2-def}$'$} %
\vec{S}_{\vec{n}} = S\left(\vec{e}_3 +m_1 \vec{e}_1+m_2\vec{e}_2\right),
\end{equation}
where $m_1$ and $m_2$ measure spin rotations moving towards the polar axis and
projected onto the $xy$--plane, respectively, see Eq.~\eqref{eq:m12-via-var}.

The spin dynamics equations of motion with an assumed $e^{-i\omega t}$ time
dependence were linearized in $m_1$ and $m_2$, leading to an eigenvalue problem
requiring numerical diagonalization. We assumed a Dirichlet boundary condition,
$m_1=m_2=0$ at the edge of the system studied. For circular systems of radius
$R$, we used a Gauss--Seidel relaxation scheme \cite{Wysin01} to calculate the
frequencies and eigenfunctions of some of the lowest eigenmodes with a single
vortex present at the system center. Before doing this, the vortex structure
was relaxed to an accurate static structure using an energy minimization
scheme. The diagonalization is partial; typically only the lowest 20 to 40
eigenstates were found, which substantially reduces the computing time needed,
and relaxes constraints on the precision of the calculations. This limited
diagonalization, however, gives only modes which have long wavelength spatial
variations, which provides for a good comparison with continuum theory.

We considered different values of $\lambda$ close to $1$. Although the
continuum limit would be better represented by using $\lambda$ very close to
$1$, this could result in a vortex radius
$r_v=\frac{a}{2}\sqrt{\lambda/(1-\lambda)}$ easily exceeding the system size
that can be treated numerically. Therefore, data were calculated using
$\lambda=0.99$, for which $r_v\approx 4.97a$. With this size of vortex length
scale, discreteness effects due to the underlying lattice should be
unimportant, and still, the vortex structure fits well within the confines of a
system with a radius as small as $R\approx 10a$, so that finite size effects
should also be negligible.

In general, a given mode has $e^{i m \chi}$ angular dependence on the azimuthal
coordinate $\chi$, where $m$ is some integer azimuthal quantum number. In the
continuum theory presented in Sec.~\ref{sec:model}, $m$ is a good quantum
number, due to rotational invariance. This symmetry is weakly broken on a
lattice, but for long-wavelength and lower frequency modes, $m$ can be
considered a good quantum number even on a lattice. (Generally, the calculated
magnon wavefunctions were sometimes found to be composed of linear combinations
of $+m$ and $-m$ components.) The numerically found modes were also
characterized by a principle quantum number $n$, being the number of nodes in
the wavefunction along the radial direction. For a mode of determined $m$ and
$n$, the scattering amplitude $\sigma$ was found by a fitting procedure applied
to the calculated eigenfunction for that mode (essentially, finding the ratio
of outgoing and incoming waves), see Ref.~\onlinecite{Wysin01} for details.

In the continuum theory, scattering was analyzed as a function of wavevector
$k$, or in terms of the dimensionless $kr_v$. For lattice calculations, the
values of $k$ cannot be chosen freely, instead, they are determined by the
actual system size. For a mode found to be oscillating at eigenfrequency
$\omega$, the wavevector magnitude $k$ associated with the mode was found by
supposing $\vec{k}=(k,0)$, and inverting the free magnon dispersion relation
for the 2D EP FM on a lattice,
\begin{equation}
\begin{split}
\omega_{\vec{k}}&=4JS\sqrt{(1-\gamma_{\vec{k}})(1-\lambda\gamma_{\vec{k}})}, \\
\gamma_{\vec{k}}&=\frac{1}{2}(\cos k_x + \cos k_y).
\end{split}
\end{equation}
Therefore, a calculation of the modes for a single lattice size gives only
specific values of $\kappa=kr_v$, one value corresponding to each mode. To get
a wider and more continuous range of data for comparison with the continuum
theory, calculations were carried out on lattices ranging in radius from
$R=15a$ to $R=40a$. By plotting results as functions of $kr_v$, for fixed $m$
but from various $n$ and $R$, the data from the different system sizes
superimposes smoothly, giving more slowly changing $kr_v$, which is more
appropriate for comparison with the continuum limit.


\subsection{Numerical results}
\label{sec:scat-disc}

\begin{figure}
\includegraphics[width=\columnwidth]{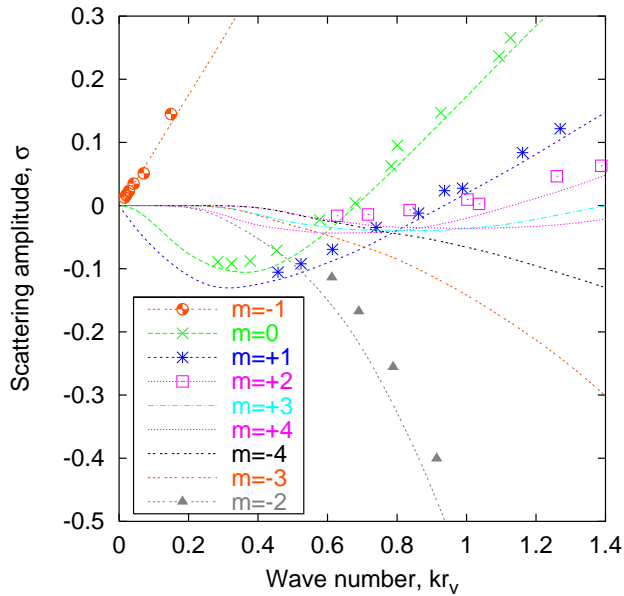}
\caption{ \label{fig:sigma-zero} %
(Color online) Scattering data for different $m$ for small wave numbers,
$kr_v<1.3$: from continuum theory (lines) and from discrete model numerical
diagonalization (symbols) in circular square lattice systems of radii $R=15,
20, 25, 30, 35, 40$.}
\end{figure}

\begin{figure}
\includegraphics[width=\columnwidth]{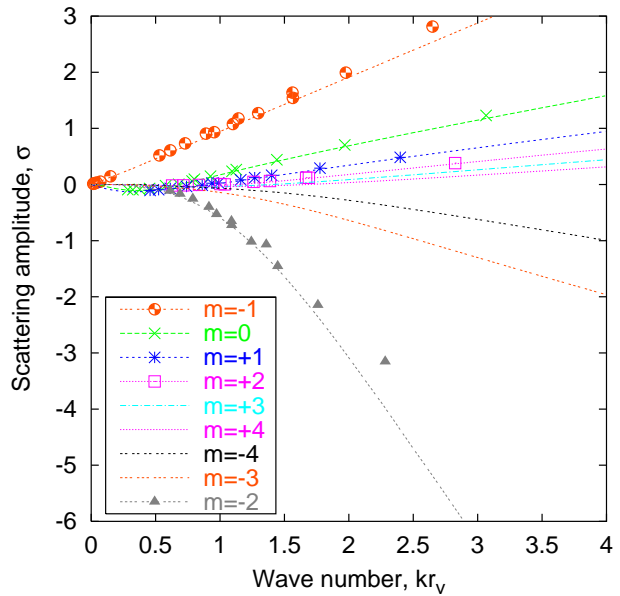}
\caption{ \label{fig:sigma} %
(Color online) Scattering data for different $m$ for a wide region of wave
numbers $k$: from continuum theory (lines) and from discrete model numerical
diagonalization (symbols).}
\end{figure}

Numerically, we have obtained the data of the vortex--magnon scattering by the
two different approaches discussed above: solving the scattering problem
\eqref{eq:PDE4num} using the shooting method for the continuum limit, and
extracting the scattering data from numerical diagonalization of finite
discrete systems. To be specific, data are presented for scattering from a
vortex with unit vorticity, $q=+1$ and positive polarization, $p =+1$. One
should note that results for vortex--magnon scattering for modes $m$ from other
vortex types, as seen in Eq.~\eqref{eq:V}, should be depend on the sign of
$qpm$. The results are the following:

For all modes the scattering amplitude $\sigma_m(k)$ tends to zero as $k\to0$.
In the long--wavelength limit the maximal scattering is related to the modes
with $m=\pm1$. Except for the mode with $m=-1$, the scattering amplitude in the
long--wavelength limit takes a negative value, see Fig.~\ref{fig:sigma-zero}.
At extremely low values of wave number $\kappa\alt 0.01$, the scattering data
contain sets of doublets for modes with opposite signs of $m$. In the
long--wavelength limit the doublet splitting appears as a small correction, but
the scattering picture changes when $k$ increases. For all modes $\sigma_m(k)$
diverges as $k\to\infty$: the scattering amplitude $\sigma_m(k\to\infty)\to
+\infty$ for all modes with $m\geq-1$, but $\sigma_m(k\to\infty)\to-\infty$ for
$m<-1$, see Fig.~\ref{fig:sigma}. Naturally, there is no real divergence; it
means that the physically observed phase shift does not tend to zero at
infinity, but to a finite value $\delta_m(k\to\infty)\to\pm\pi/2$. The
scattering data are presented in Figs.~\ref{fig:sigma-zero}, \ref{fig:sigma}.
Comparison with the results of exact diagonalization on finite systems shows
very good agreement between the two approaches.


\section{Scattering problem: analytical description} %
\label{sec:scat-analyt}


\subsection{Scattering at long wavelength} %
\label{sec:k<<1}

In order to analyze the scattering problem analytically in the long--wavelength
limit, we start from the zero--frequency solutions, when $\varOmega=0$. First
note that for the special cases $m=0,\pm1$ there exist so--called
\emph{half--bound states}. Recall that a zero--frequency solution of the
Schr\"{o}dinger--like equation is called a half bound state if its wave
function is finite, but does not decay fast enough at infinity to be square
integrable. We will refer to such modes as \emph{half--local modes}. These
modes correspond to the translational ($m=\pm1$) and rotational ($m=0$)
symmetry of an infinite system, they have an exact analytical form:
\begin{equation} \label{eq:zero-mode-m<=1}
\tilde{u}_m^{(0)}=\frac{q\cdot\sin\theta_0}{\rho^{|m|}},\;
\tilde{v}_m^{(0)}=m\cdot\theta_0^\prime,\qquad m=0,\pm1. \end{equation} Unlike
the case of half--local modes with $m=0,\pm1$, all other zero--frequency
solutions are nonlocal, and we are not able to construct exact expressions for
them, but only the asymptotes for $\rho\gg1$:
\begin{equation} \label{eq:zero-mode-m>1}
\tilde{u}_m^{(0)}\propto \rho^{|m|},\qquad \tilde{v}_m^{(0)}\propto
\frac{e^{-\varkappa\rho}}{\sqrt{\rho}}.
\end{equation}
Nevertheless, we will see that the knowledge of asymptotic solutions like
\eqref{eq:zero-mode-m>1} will be enough to reconstruct the $\kappa$--dependence
of the scattering amplitude. In order to solve the scattering problem in the
long wavelength limit we apply a special perturbation theory, proposed in
Ref.~\onlinecite{Ivanov98,Ivanov00} for the modes with $m=\pm1,0$, and
extending it for all values of $m$. We construct the asymptotes of such a
solution for a small but finite frequency by making the ansatz
\begin{subequations} \label{eq:alpha&beta}
\begin{eqnarray} \label{eq:alpha}
\tilde{v}(\rho) &=& \tilde{v}_0(\rho)\cdot\Bigl[1 + \kappa\alpha_1(\rho)
+\kappa^2\alpha_2(\rho)\Bigr],\\
\label{eq:beta} \tilde{u}(\rho) &=& \tilde{u}_0(\rho)\cdot\Bigl[1 +
\kappa\beta_1(\rho) + \kappa^2\beta_2(\rho)\Bigr],
\end{eqnarray}
\end{subequations}
here $\alpha_1$, $\beta_1$ and $\alpha_2$, $\beta_2$ are first and second order
corrections to the zeroth--solutions, respectively. Let us insert this ansatz
into the set of Eqs.~\eqref{eq:PDE4num}, multiply from the left with $\rho\cdot
\bm{\bigl<\widetilde{\varPsi}\bigr|}$ without integrating; then one obtains
equations for the first and second order corrections:
\begin{widetext}
\begin{equation} \label{eq:prime-of-prime}
\begin{split}
&\Bigl[\rho\cdot\left(\alpha_k^\prime\cdot \tilde{v}_0^2 + \beta_k^\prime\cdot
\tilde{u}_0^2\right)\Bigr]^\prime = \Phi_k(\rho),\qquad k=1,2,\qquad
\Phi_1(\rho) = 2\rho\Bigl\{V\left( \tilde{u}_0^2- \tilde{v}_0^2\right) +
2(W-1/2)\tilde{u}_0\tilde{v}_0\Bigr\},\\
&\Phi_2(\rho) = \rho\biggl\{-\tilde{u}_0^2+\tilde{v}_0^2/2+ 2(W-1/2) \left(
\tilde{u}_0^2- \tilde{v}_0^2\right)+2V\left( \tilde{u}_0^2\beta_1 -
\tilde{v}_0^2\alpha_1\right) +
2\tilde{u}_0\tilde{v}_0\bigl[(W-1/2)(\alpha_1+\beta_1)-2V \Bigr]\biggr\}.
\end{split}
\end{equation}
We are interested in the corrections $\beta_k$, which will give us a
possibility to calculate the scattering amplitude. The formal solution of these
equations can be written as
\begin{equation} \label{eq:beta-formal}
\begin{split}
\beta_k(\rho) &= \beta(0) + \int_0^\rho \frac{\alpha_k^\prime(\eta)
\tilde{v}_0^2(\eta) \text{d}\eta}{\tilde{u}_0^2(\eta)} + \int_0^\rho
\frac{\text{d}\eta}{\eta \tilde{u}_0^2(\eta)} \int_0^\eta
\Phi_k(\xi)\text{d}\xi.
\end{split}
\end{equation}
\end{widetext}
Let us calculate the first order correction $\beta_1$. It is easy to see that
the second RHS--term has an exponential decay as $\rho\to\infty$, while the
third one has a slow algebraic decay only. Thus, far from the vortex core we
have simply
\begin{equation} \label{eq:beta1}
\beta_1(\rho) \simeq \text{const} -\frac{\rho^{-2|m|}}{2|m|} \int_0^\infty
\Phi_1(\xi)\text{d}\xi,
\end{equation}
valid in the region $\rho \gg 1$.

To calculate the second order correction, $\beta_2$, let us note that the last
RHS--term of Eq.~\eqref{eq:beta-formal} is divergent for $\rho\to\infty$, while
the integral with $\alpha_2$ has an exponential decay, like the first order
correction. To derive the divergent inner integral in
Eq.~\eqref{eq:beta-formal} we add and subtract the function
\begin{equation} \label{eq:Phi_2^0}
\Phi_2^{(0)}(\xi) = -\frac{\left[\xi^2\tilde{u}_0^2\right]^\prime}{2(|m|+1)} -
\frac{\left[\sin^2\theta_0\tilde{u}_0^2\right]^\prime}{2|m|}.
\end{equation}
Then we arrive at an approximation for $\beta_2(\rho)$ in the important region
$\rho\gg1$:
\begin{equation} \label{eq:beta2}
\begin{split}
\beta_2(\rho) &\simeq \text{const} -\frac{\rho^2}{4(|m|+1)} - \frac{\ln\rho}{2|m|}\\
&- \frac{\rho^{-2|m|}}{2|m|} \int_0^\infty
\left[\Phi_2(\xi)-\Phi_2^{(0)}(\xi)\right]\text{d}\xi.
\end{split}
\end{equation}

Now we are in position to compare the magnon amplitude
$\tilde{u}_m=\tilde{u}_0(1+\kappa\beta_1+\kappa^2\beta_2)$ with the scattering
approach in order to extract the information about the scattering amplitude
$\sigma_m(\kappa)$. To describe the scattering problem in the long--wavelength
approximation we rewrite the differential problem
\eqref{eq:PDE4tilde-Psi+tilde-H} for large distances $\kappa\rho\gg1$, only
considering the terms with $\kappa^2$. In this scattering approach the
oscillating function $\tilde{u}_m$ satisfies an equation
\begin{equation} \label{eq:eq4u-small-k}
\left(\nabla_\rho^2+\kappa^2-\frac{\nu^2}{\rho^2}\right)\tilde{u}_m=0,\qquad
\nu^2=m^2-\kappa^2.
\end{equation}
The solution of this equation can be written as
\begin{equation} \label{eq:u-small-k}
\begin{split}
\tilde{u}_m(\rho)&\propto J_{|\nu|}(\kappa\rho)+\tilde{\sigma}_\nu(\kappa)Y_{|\nu|}(\kappa\rho)\\
&\propto\frac{1}{\sqrt{\rho}}\cos\left(\kappa\rho-\frac{|\nu|\pi}{2}-\frac{\pi}{4}
+\tilde{\delta}_m\right),
\end{split}
\end{equation}
where the index of the Bessel and the Neumann function is noninteger. It
results in a value of $\tilde{\delta}_m$ which differs from the real scattering
phase shift $\delta_m$. Using asymptotic expansions \eqref{eq:u4x>>1} and
\eqref{eq:u-small-k}, the desired relation between the phase shift and
$\tilde{\delta}_m$ can be written as
\begin{equation}
\label{eq:delta-via-tilde-delta}
\delta_m(\kappa)=\tilde{\delta}_\nu(\kappa)+\frac{|m|-|\nu|}{2}\pi.
\end{equation}
In the lowest order approximation in $\kappa$, the corresponding relation for
the scattering amplitudes has the form:
\begin{equation} \label{eq:sigma-via-tilde-sigma}
\sigma_m(\kappa)=\tilde{\sigma}_\nu(\kappa) - \frac{\pi\kappa^2}{4|m|}.
\end{equation}
To compare the scattering solution \eqref{eq:u-small-k} with the result of the
perturbation theory we can expand the cylindrical functions in powers of the
small quantity $|\nu|-|m|$ and represent through the cylindrical functions of
integer order $|m|$, as done in Ref.~\onlinecite{Ivanov02}. After that in the
region $\kappa\rho\ll1$ we are able to use the asymptotes of the cylindrical
functions at the origin; we arrive at the formula
\begin{equation} \label{eq:u-from-scat}
\begin{split}
\tilde{u}_m(\rho) &\simeq \rho^{|m|}\Biggl\{
1-\frac{\kappa^2\rho^2}{4(|m|+1)}-\frac{\kappa^2}{2|m|}\ln\left(\frac{\kappa\rho}{2}\right)
\\ &-\sigma_m\cdot\frac{(|m|!)^2}{\pi|m|}\left(\frac{2}{\kappa\rho}\right)^{2|m|}
\left[1 - \frac{\kappa^2}{4|m|}S_m\right]\Biggr\},\\
S_m &=\gamma+|m|\sum_{n=1}^{|m|-1}\frac{1}{n(|m|-n)},
\end{split}
\end{equation}
where $\gamma$ is Euler's constant.

Comparing this expression with the perturbation theory results [see Eqs.
\eqref{eq:beta}, \eqref{eq:beta1}, \eqref{eq:beta2}], in the region
$1\ll\rho\ll1/\kappa$, where both are valid, we can restore the general
dependence of the scattering amplitude in the long--wavelength approximation:
\begin{subequations}  \label{eq:sigma4k<<1+A+B}
\begin{align} \label{eq:sigma4k<<1}
\sigma_m(k) &=-\mathcal{A}_m\left(\frac{\kappa}{2}\right)^{2|m|} +
m\mathcal{B}_m\left(\frac{\kappa}{2}\right)^{2|m|+1},\\
\label{eq:sigma4k<<1(A)}
\mathcal{A}_m&=\frac{2\pi|m|}{S_n(|m|!)^2}\int_0^\infty\left[\Phi_2(\xi)-
\Phi_2^{(0)}(\xi)\right]\text{d}\xi,\\
\label{eq:sigma4k<<1(B)}
\mathcal{B}_m&=-\frac{\pi}{m(|m|!)^2}\int_0^\infty\Phi_1(\xi)\text{d}\xi.
\end{align}
\end{subequations}
Eq.~\eqref{eq:sigma4k<<1} solves the scattering problem except for factors
$\mathcal{A}_m$ and $\mathcal{B}_m$. These factors can be found by the
numerical integration of Eqs. \eqref{eq:sigma4k<<1(A)} and
\eqref{eq:sigma4k<<1(B)}, using numerical data for $\tilde{u}_0$ and
$\tilde{v}_0$. Thus, solving the equations for zero--modes once, we compute the
whole dependence $\sigma_m(\kappa)$. Nevertheless, in order to discuss the
analytical behavior let us note that for sufficiently large values of $|m|$, we
can limit ourselves to the contribution of the term with $\tilde{u}_0^2$ in the
function $\Phi_1$ and the term with $\tilde{u}_0^2\beta_1$ in the function
$\Phi_2$, see Eq.~\eqref{eq:prime-of-prime}. To calculate the integrals we need
to have more information about the zero--modes. At small distances $\rho\ll1$
the isotropic (exchange) approximation works correctly, which leads to the
following solutions:\cite{Ivanov95g}
\begin{equation*}
\tilde{u}_0\propto\rho^{|m|}\sin\theta_0,\quad\tilde{v}_0\propto
\rho^{|m|+1}\theta_0^\prime.
\end{equation*}
Such solutions have the correct asymptotic behavior at infinity and at the
origin. Our numerical calculations justify the correctness of these assumptions
for $m\gg1$; as a result we obtain analytical estimates for these factors:
\begin{equation} \label{eq:coef4k<<1}
\mathcal{A}_m \approx \frac{18(2|m|-1)!}{S_n((|m|-1)!)^2},\quad \mathcal{B}_m
\approx \frac{4\pi(2|m|-1)!}{(|m|!)^2}.
\end{equation}
To compare the scattering results for different modes, we write explicitly the
asymptotic expressions for all modes, taking into account
\eqref{eq:sigma4k<<1}, and asymptotes for half--local modes from
Refs.~\onlinecite{Ivanov98,Ivanov00}
\begin{subequations} \label{eq:sigma-as-all4k<<1} \begin{align}
\label{eq:sigma-as-all4m=0} %
\sigma_{m=0}(\kappa)&\approx -\frac
\pi2\kappa^2\ln(1/\kappa),\\ \label{eq:sigma-as-all4m=pm1} %
\sigma_{m=\pm1}(\kappa) &\approx \mp\frac{\pi\kappa}{4},\\
\label{eq:sigma-as-all4|m|>1} %
\sigma_{m\neq0,\pm1}(\kappa) &\approx
-\mathcal{A}_m\left(\frac{\kappa}{2}\right)^{2|m|} +
m\mathcal{B}_m\left(\frac{\kappa}{2}\right)^{2|m|+1}. \end{align}
\end{subequations}

In the main approximation in $\kappa$ the scattering picture contains doublets
for modes with opposite signs of $m$ for the modes with $|m|>1$.
The splitting of the doublets (the last term in
Eq.~\eqref{eq:sigma-as-all4|m|>1}) appears in the next order in $\kappa$. The
splitting of the doublets for the magnon modes on the vortex background with
given $m=\pm n$ was mentioned in the earliest papers on vortex--magnon
scattering;\cite{Wysin95,Wysin96} but it was explained, in fact, only for
$m=\pm1$. \cite{Ivanov98} Our considerations on the basis of
Eq.~\eqref{eq:Gen-Schroedinger} show that the splitting of the scattering data
is the direct analogue of the Zeeman effect for electron states splitting in an
external magnetic field. To follow this analogy one can rewrite the splitting
constant $ \mathcal{B}_m$ in the form:
\begin{equation*}
\mathcal{B}_m\propto \int_0^\infty\! \text{d}\xi\;
\xi^{2|m|}\left(\vec{A}(\xi)\cdot \vec{e}_\chi\right),
\end{equation*}
hence the splitting appears only in the effective magnetic field, which is
described by the vector potential $\vec{A}$.

Using scattering results \eqref{eq:sigma-as-all4k<<1} one can solve the
scattering problem for a plane spin wave in the form \eqref{eq:G4plane-wave}.
In the long--wavelength limit the maximum scattering is related to the
translation modes with $m=\pm1$, which gives the scattering function
\eqref{eq:G4plane-wave2} in the form
\begin{equation} \label{eq:ScatF4k<<1-result}
\mathcal{F}(\chi) = \sqrt{\frac{\pi\kappa}{2}}e^{3{i} \pi/4}\sin\chi.
\end{equation}
In this approximation the scattering is anisotropic, and the total scattering
cross section is $\varrho = \pi^2\kappa/4$. To explain the origin of the
anisotropic scattering, let us mention that the plane spin wave makes a spin
flux, which influences the vortex as a whole, trying to move it by exciting
translational modes. It is well--known that the vortex dynamics appears in the
gradient of a magnetization field like the magnon flux. \cite{Nikiforov83} The
dynamics of the vortex has a gyroscopical behavior (see for the review
\onlinecite{Mertens00}): acting along the $x$--axis, the spin wave causes the
translational motion of the vortex along the $y$--axis, which results in
\eqref{eq:ScatF4k<<1-result}.


\subsection{Scattering problem for short wavelength} %
\label{sec:k>>1}

\begin{figure}
\includegraphics[width=\columnwidth]{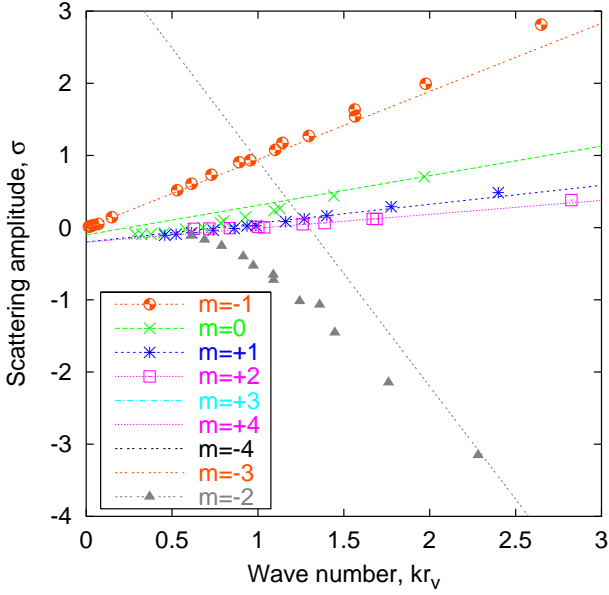}
\caption{ \label{fig:sigma-infty} %
(Color online) Scattering data for different $m$ for the short--wavelength
limit: from asymptotes \eqref{eq:sigma-via-C} of the continuum theory (lines)
and from discrete model numerical diagonalization (symbols)}
\end{figure}

For large $k$, in the main approximation to lowest order in $1/\kappa$, the
scattering problem \eqref{eq:PDE4num} can be rewritten in the form:
\begin{subequations} \label{eq:PDE4num-WKB}
\begin{align} \label{eq:PDE4num-WKB(1)} %
\widetilde{\textsf{H}} \bm{\bigl|\widetilde{\varPsi}\bigr>} &= -\kappa^2
\bm{\bigl|\widetilde{\varPsi}\bigr>},\\
\label{eq:PDE4num-WKB(2)} %
\bm{\bigl|\widetilde{\varPsi}\bigr>} & \sim
\begin{Vmatrix}\epsilon_m\cdot\rho^{|m+1|},\\ %
\rho^{|m-1|}\end{Vmatrix},\qquad \text{when}\;\rho\ll1\\
\label{eq:PDE4num-WKB(3)} %
\bm{\bigl|\widetilde{\varPsi}\bigr>} & \sim
\begin{Vmatrix}
J_{|m|}(\kappa\rho)+\sigma_m\cdot Y_{|m|}(\kappa\rho) \\ %
K_{|m|}(\kappa\rho)
\end{Vmatrix},\qquad\text{when}\;\rho\gg1.
\end{align}
\end{subequations}
We see that the functions $\tilde{u}$ and $\tilde{v}$ have independent
asymptotes at the origin \eqref{eq:PDE4num-WKB(2)} and at infinity
\eqref{eq:PDE4num-WKB(3)}. It means that  the role of the ``coupling
potential'' $W$ in the scattering problem \eqref{eq:PDE4num-WKB} is unimportant
here. Therefore one can neglect the ``coupling potential'' and formulate the
scattering problem for the master function $\tilde{u}$ only:
\begin{equation} \label{eq:Scroedinger4Um} %
\left[-\nabla_\rho^2+\mathcal{U}_m(\rho)\right] \tilde{u}_m = \kappa^2
\tilde{u}_m\;,
\end{equation}
where the partial potential is
\begin{equation} \label{eq:Um} %
\mathcal{U}_m(\rho)= \mathcal{U}_0(\rho) + V(\rho) = U(\rho)-\frac12+\left[
\left(\vec{A}(\rho)\cdot\vec{e}_\chi\right)-\frac{m}{\rho}\right]^2.
\end{equation}

It is natural to suppose that the WKB--approximation is valid for this case. We
use the WKB--method in the form proposed earlier for the description of the
scattering for isotropic 2D magnets \cite{Ivanov99}, and generalized after that
for any singular potentials. \cite{Sheka02} We start from the effective 1D
Schr\"{o}dinger equation for the radial function $\tilde{u}_m(\rho) =
\psi_m(\rho)/\sqrt{\rho}$, which yields
\begin{equation} \label{eq:U-ef}
\begin{split}
&\left[-\frac{d^2}{d\rho^2}+\mathcal{U}_{\text{eff}}(\rho)\right] \psi_m =
\kappa^2
\psi_m\;,\\
&\mathcal{U}_{\text{eff}}(\rho)= \mathcal{U}_m(\rho)-\frac{1}{4\rho^2}\;.
\end{split}
\end{equation}
The WKB--solution of Eq.~\eqref{eq:U-ef}, i.e. the 1D wave function
$\psi_m^{WKB}$, leads to the following form of the partial wave
\begin{equation} \label{eq:WKB}
\tilde{u}_m^{WKB} = \frac{\psi_m^{WKB}}{\sqrt{\rho}} \propto
\frac{1}{\sqrt{\rho\cdot\mathcal{P}(\rho)}}\cos\left(\chi_0 +
\int_{\rho_0}^\rho \mathcal{P}(\rho')d\rho'\right),
\end{equation}
where $\mathcal{P}=\sqrt{\kappa^2-\mathcal{U}_{\text{eff}}}$. Analysis shows
that Eq.~\eqref{eq:WKB} is valid for $\rho>a$, where $a$ is the turning point.
The value of $a$ corresponds to the condition $\mathcal{P}(a)=0$, which results
in $a\sim|m|/\kappa\ll 1$. We assume that the parameter $\rho_0$ satisfies the
condition $a\ll \rho_0 \ll 1$.

On the other hand, at small distances $\rho\ll1$, the partial potential
$\mathcal{U}_m$ has the asymptotic form
\begin{equation} \label{eq:U-as0}
\begin{split}
&\mathcal{U}_m \sim \frac{\nu^2}{\rho^2},\\ %
\nu&= m-\lim_{\rho\to0}\left[\rho\left(\vec{A}(\rho)\cdot\vec{e}_\chi\right)
\right] = m+qp,
\end{split}
\end{equation}
therefore one can construct asymptotically exact solutions (recall that we
suppose $q=p=1$)
\begin{equation} \label{eq:psi-WKB0}
\tilde{u}_m \propto J_{|m+1|}(\kappa \rho),\qquad\text{when $\rho\ll 1$}.
\end{equation}
For $\kappa\gg |m|$ there is a wide range of values of $\rho$, namely
\begin{equation} \label{eq:wide-range}
|m|/k\ll \rho\ll 1,
\end{equation}
where we can use the asymptotic expression \cite{Ivanov99} for the Bessel
function \eqref{eq:psi-WKB0} in the limit $k\rho\gg|m|$:
\begin{equation} \label{eq:psi-WKB4rho>>1}
\tilde{u}_m \propto \frac{1}{\sqrt{\rho}}\cos\Bigl(\kappa \rho -
\frac{|m+1|\pi}{2} -\frac\pi4+\frac{4|m+1|^2-1}{8\kappa \rho}\Bigr).
\end{equation}
In the range of Eq.~\eqref{eq:wide-range} the solutions \eqref{eq:WKB} and
\eqref{eq:psi-WKB4rho>>1} coincide due to the overlap of the entire range of
parameters, so one can derive the phase $\chi_0$ in the WKB--solution
\eqref{eq:WKB},
\begin{equation*}
\chi_0 = \kappa \rho_0 - \frac{|m+1|\pi}{2}
-\frac\pi4+\frac{4|m+1|^2-1}{8\kappa \rho_0}\;.
\end{equation*}
Therefore, we are able to calculate the short--wavelength asymptotic expression
for the scattered wave phase shift by the asymptotic expansion of the
WKB--solution \eqref{eq:WKB}:
\begin{equation} \label{eq:delta-via-P}
\begin{split}
\delta_m(\kappa) = \lim_{\rho\to\infty}&
\Biggl(\int_{\rho_0}^\rho\mathcal{P}(\rho')d\rho' +\chi_0 - \kappa \rho\\
&+\frac{|m|\pi}{2}+\frac{\pi}{4}-\frac{4m^2-1}{8\kappa \rho}\Biggr).
\end{split}
\end{equation}
Under assumed conditions $k\rho\gg1$, the WKB--integral in
\eqref{eq:delta-via-P} can be calculated in the leading approximation in
$1/k\rho$,
\begin{equation*}
\int_{\rho_0}^\rho\mathcal{P}(\rho')d\rho' \approx
\kappa(\rho-\rho_0)-\frac{1}{2\kappa} \int_{\rho_0}^\rho
\mathcal{U}_{\text{eff}}(\rho')d\rho'.
\end{equation*}
As a result, the scattering phase shift for large wave numbers, $\kappa\gg 1$,
has the form
\begin{equation*}
\begin{split}
\delta_m(\kappa) &= \delta_m(\infty) - \frac{1}{2\kappa}\int_0^\infty
\Bigl[\mathcal{U}_m(\rho)- \frac{\nu^2}{\rho^2}\Bigr] d\rho,
\end{split}
\end{equation*}
with the limiting value
\begin{align} \label{eq:delta-infty}
\delta_m(\infty)&= -\frac\pi2\cdot\left(|\nu|-|m|\right) = -\frac\pi2\cdot
\sign_+(m),\\
\nonumber %
\sign_+(m) &=
\begin{cases}
1,& m\geq0,\\-1,&m<0.
\end{cases}
\end{align}
Calculating the integral we obtain the phase shift in the form:
\begin{eqnarray} \label{eq:delta-via-D}
&&\delta_m(\kappa) = -\frac{\pi}{2}\cdot \sign_+(m)
+\frac{\mathcal{D}_1+m\mathcal{D}_2}{\kappa}
\\
\nonumber %
&&\mathcal{D}_1=\frac14\int_0^\infty \left\{\frac{3\sin^2\theta_0}{\rho^2}
+3\cos^2\theta_0+\left(\theta_0^\prime\right)^2\right\}d\rho\approx 2.44,\\
\nonumber %
&&\mathcal{D}_2=\int_0^\infty \frac{1-\cos\theta_0}{\rho^2}d\rho\approx 1.38.
\end{eqnarray}
The corresponding amplitude of the vortex--magnon scattering is
\begin{equation} \label{eq:sigma-via-C}
\sigma_m(\kappa) = \frac{\kappa}{\mathcal{D}_1+m\mathcal{D}_2}.
\end{equation}
This linear divergence is well--pronounced in the numerical results, see
Fig.~\ref{fig:sigma-infty}. To understand the origin of this divergence let us
go back to the Eq.~\eqref{eq:delta-infty}. One can see that the scattering
phase shift at $k\to\infty$ does not vanish for potentials with inverse square
singularity at the origin, with $\nu\neq m$, see Eq.~\eqref{eq:U-as0}. This is
possible only in the magnetic field, which has singular behavior like
$|\vec{A}| \sim1/\rho$.

Let us look for the consequences of this unusual behavior of the scattering,
$\sigma_m\to\pm\infty$. We consider the scattering problem for a plane spin
wave in the form \eqref{eq:G4plane-wave}. In the short--wavelength limit the
WKB results for the phase shift \eqref{eq:delta-via-D} are available. One can
see that the scattering function \eqref{eq:G4plane-wave2} tends to zero very
quickly for large wave numbers, $\mathcal{F}(\chi) =
\mathcal{O}(\kappa^{-5/2})$, so there is no real divergence or singularity for
a physically observable quantity such as the total scattering function
$\mathcal{F}$ at large energies.


\subsection{Levinson theorem} %
\label{sec:Levinson}

\begin{figure}
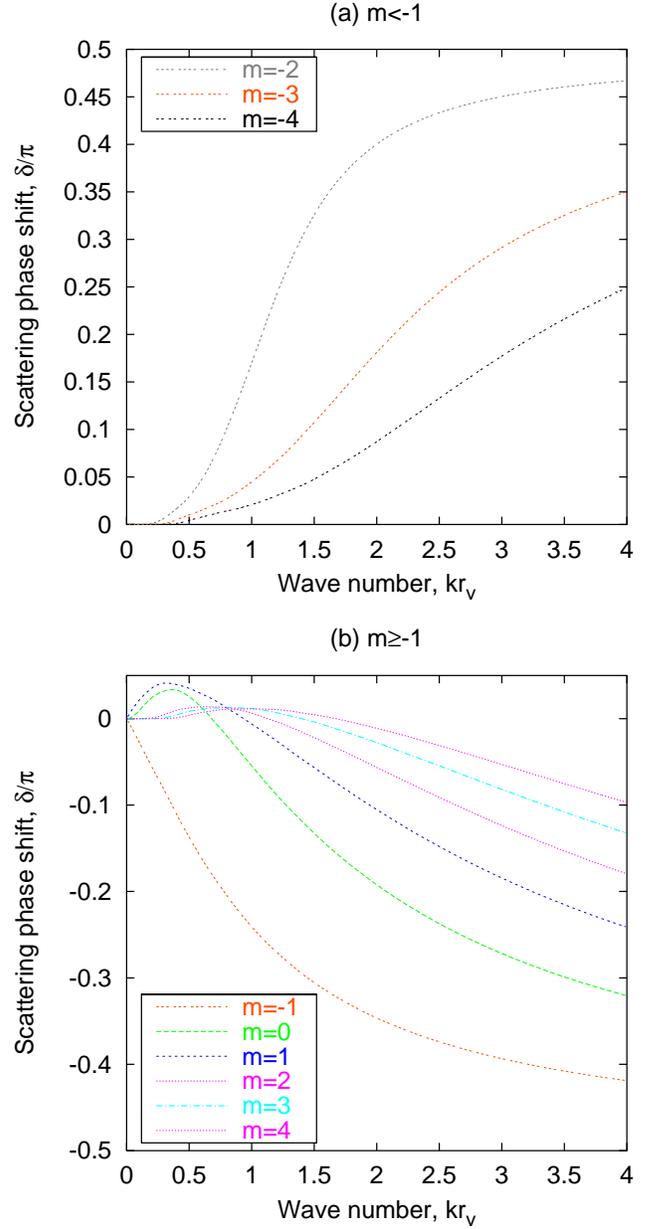

\includegraphics[width=3.5in]{\myfig{4a}}
\includegraphics[width=3.5in]{\myfig{4b}}
\caption{ \label{fig:delta} %
(Color online) Scattering phase shifts for different $m$. Numerical results
from the continuum theory.}
\end{figure}

Now we can compare the scattering results in the long-- and short--wavelength
limits. The scattering is absent for the limit $k\to0$. However, the scattering
amplitude has a linear divergence $\sigma\propto k$ for sufficiently large wave
numbers, see Eq.~\eqref{eq:sigma-via-C}. All these results were verified by the
numerical calculations for continuum limit and for finite sized discrete
lattice systems, see Figs. \ref{fig:sigma-zero}, \ref{fig:sigma},
\ref{fig:sigma-infty}. According to our analytical calculations, see
Eq.~\eqref{eq:delta-infty}, the phase shift for the short--wavelength limit
tends to the finite value $\delta_m(\infty) = -\sign_+(m)\cdot \pi/2$. This
result corresponds to the numerical data, see Fig.~\ref{fig:delta}, except for
the mode with $m=-1$, where the numerical data gives
$\delta_{m=-1}(\infty)=-\pi/2$. However we need to note that the phase shift is
determined with respect to $\pi$, in this sense values $\delta_m=\pi/2$ and
$\delta_m=-\pi/2$ are identical. What is physically important is how
$\delta_m(\kappa)$ changes from small to large $\kappa$. According to our
numerical results, the total phase shift can be described by the formula
\begin{equation} \label{eq:Levinson}
\delta_m(0) - \delta_m(\infty) =
  \begin{cases}
\frac\pi2\cdot\sign_+(m), & m\neq-1, \\
\frac\pi2, & m=-1.
  \end{cases}
\end{equation}

It is well--known that the total phase shift is related to the number of bound
states $N_m^{\text{b}}$ according to the Levinson theorem from the scattering
problem for a spinless quantum mechanical particle without a magnetic field.
This theorem was originally proved by Levinson for the 3D case, see the review
\onlinecite{Newton82}. The two-dimensional version of the Levinson theorem
reads \cite{Bolle86,Lin97,Dong98}
\begin{subequations} \label{eq:Levinson-2D}
\begin{equation} \label{eq:Levinson-2D(1)}
\delta_m(0) - \delta_m(\infty) = \pi\cdot N_m^{\text{b}}.
\end{equation}
If there exist half--bound states (see notations in the Sec.~\ref{sec:k<<1})
for the $p$--wave ($m=1$), this is modified to \cite{Bolle86,Dong98}
\begin{equation} \label{eq:Levinson-2D(2)}
\delta_1(0) - \delta_1(\infty) = \pi\cdot N_1^{\text{b}}+\pi.
\end{equation}
\end{subequations}
For the 2D EP FM, the scattering picture is much more complicated. First, we
have no standard Schr\"{o}dinger equation, but the generalized one
\eqref{eq:Gen-Schroedinger}. This becomes apparent at most in the threshold
behavior for the half--bound states, and the contribution of the half--bound
states in the form \eqref{eq:Levinson-2D(2)} may be not adequate, see below.
Second, because of the role of the effective magnetic field, there appears an
$m$--dependent potential: the symmetry $\delta_m(\kappa)=\delta_{-m}(\kappa)$
is broken, so it is not enough to take into account partial waves with $m\ge0$
only. As a result Levinson's relation \eqref{eq:Levinson-2D(1)} has a different
form for the opposite signs of $m$.

Thus, except for the case of half--bound states one can hope that the Levinson
theorem is adequate. However we see that the total phase shift
\eqref{eq:Levinson} contradicts the Levinson theorem in the form
\eqref{eq:Levinson-2D}. The reason is that the partial potential
$\mathcal{U}_m$ in the Schr\"{o}dinger equation \eqref{eq:Scroedinger4Um} has
an inverse square singularity at the origin, $\mathcal{U}_m\sim\nu^2/\rho^2$,
where $\nu=m+qp$, see \eqref{eq:U-as0}. Such a situation changes the statement
of the Levinson theorem. As we have proved recently in
Ref.~\onlinecite{Sheka03}, the generalized Levinson theorem for the
Schr\"{o}dinger--like equation for potentials with such singularities has form:
\begin{equation} \label{eq:Levinson-2D-gen}
\delta_m(0) - \delta_m(\infty) = \pi\cdot N_m^{\text{b}}
+\frac{\pi}{2}\left(|\nu|-|m| \right).
\end{equation}
An additional $\pi$ can appear on the RHS of this equation, if the half--bound
states exist for the $p$--wave ($|m|=1$), see \eqref{eq:Levinson-2D(2)}. To
explain the meaning of the extra term $(\pi/2)\cdot(|\nu|-|m|)$ in the
generalized Levinson theorem \eqref{eq:Levinson-2D-gen}, recall that in the
partial wave method the scattering data are classified by the azimuthal quantum
number $m$, which is the strength of the centrifugal potential. In the presence
of a partial potential with an inverse square singularity at the origin such as
$\mathcal{U}_m\sim \nu^2/\rho^2$, the effective singularity strength is shifted
by the value $|\nu|-|m|$, which results in a change in the short--wavelength
scattering phase shift by $(\pi/2)\cdot(|m|-|\nu|)$.

Let us compare the predictions of the generalized Levinson theorem
\eqref{eq:Levinson-2D-gen}, which is suitable for the Schr\"{o}dinger--like
equation, with our results for the vortex--magnon scattering problem in the 2D
EP FM, which can be described by the generalized Schr\"{o}dinger equation
\eqref{eq:Gen-Schroedinger}. In our case the singular potential is caused by
the specific singular magnetic field at the origin, $|\vec{A}|\sim 1/\rho$,
which results in $\nu=|m+1|$. The system has no bound states,
$N_m^{\text{b}}=0$, therefore Eq.~\eqref{eq:Levinson-2D-gen} takes the form:
\begin{equation} \label{eq:Levinson-2D-gen1} \delta_m(0) - \delta_m(\infty) =
\frac\pi2\cdot\sign_+m. \end{equation} Our numerical results
\eqref{eq:Levinson} correspond to this formula for all modes with $m\neq-1$.
The cause is the influence of the half--bound states. By comparison of
\eqref{eq:Levinson-2D-gen1} and \eqref{eq:Levinson}, one can adapt the
generalized Levinson theorem for this case. It reads
\begin{equation} \label{eq:Levinson-final}
\begin{split}
&\delta_m(0) - \delta_m(\infty)\\ &=
\begin{cases}
\pi\cdot N_m^{\text{b}}+\frac{\pi}{2}\left(|\nu|-|m| \right), &
\text{when $m\neq-1$}\\
\pi\cdot N_m^{\text{b}}+\pi+\frac{\pi}{2}\left(|\nu|-|m| \right), & \text{when
$m=-1$}.
\end{cases}
\end{split}
\end{equation}
Let us compare this result with Eq.~\eqref{eq:Levinson-2D}. An extra $\pi$,
which appears for the mode $m=-1$, is connected with the half--bound states,
see Eq.~\eqref{eq:Levinson-2D(2)}. To explain the situation, let us stress
again that our scattering problem is formulated not for the standard
Schr\"{o}dinger equation. However, the problem has a symmetry such that one
eigenfunction becomes a master function, while the other is a slave. This makes
it possible to use the main features of the standard quantum mechanical
scattering theory. The appearance of the half--bound states is connected with
the symmetry of the whole system, and both of the eigenfunctions are important.
In the system there are three half--local modes, see
Eq.~\eqref{eq:zero-mode-m<=1}. According to Eq.~\eqref{eq:Levinson-final} only
one of the half--bound modes, namely, the mode with $m=-1$, gives an extra
$\pi$ to the Levinson's relation. More generally, this extra contribution
corresponds to the half--bound mode with $m=-qp$, see Eq.~\eqref{eq:U-as0}.
This result cannot be explained in the framework of the Levinson theorem for
the standard Schr\"{o}dinger equation, where both half--bound states with
$m=+1$, and $m=-1$ should make contributions to the Levinson's relation. The
corresponding analogue of the Levinson theorem for the generalized
Schr\"{o}dinger equation \eqref{eq:Gen-Schroedinger} takes into account the
contribution of the half--bound state for only \emph{one} value of $m$, namely,
for $m=-qp$.


\section{Conclusion}
\label{sec:concl}

We have presented a detailed study of vortex--magnon interactions in the 2D EP
ferromagnet, having described this process by a ``generalized'' Schr\"{o}dinger
equation. The main features of the magnon scattering are connected with the
special role of the effective magnetic field, which is created by the vortex.
This effective field acts on magnons in the same way as a magnetic field
influences an electron, leading to the appearance of the Lorentz force and the
Zeeman splitting of the magnon states with opposite values of the azimuthal
numbers $m$. The singular behavior of the effective magnetic field at the
origin causes a divergence of the scattering amplitudes for all the partial
waves; we have confirmed this study by a generalized version of the Levinson
theorem for potentials with inverse square singularities.

Our investigations can be applied to the description of the internal dynamics
of vortex state magnetic dots; the theory of the vortex--magnon scattering
developed here could be a good guide for the study of the normal modes in
vortex--state magnetic dots. It is clear that the EP FM cannot correspond
quantitatively to the case of vortex--state magnetic dots, where the anisotropy
is negligible and the static vortex structure is stabilized by magnetic--dipole
interactions. We did not consider this type of interaction in this paper, as it
is difficult to account for. Nevertheless, we believe that the main features of
the problem studied above are generic. For example, an effective magnetic field
exists due to the topological properties of the vortex only. Furthermore, we
expect the appearance of modes with anomalously small frequencies, e.g., the
mode of the translational oscillations of the vortex center. The nonzero
frequency of this mode is caused by the interaction with the boundary only.
Additionally, the splitting of the doublets for modes with opposite $m$ should
appear due to the role of the effective magnetic field.

\begin{acknowledgments}
D.Sh. thanks the University of Bayreuth, where part of this work was performed,
for kind hospitality and acknowledges support by the European Graduate School
``Non--equilibrium phenomena and phase transitions in complex systems'', and
the DLR Project No. UKR-02-011.
\end{acknowledgments}

%
%

%
%
\end{document}